\documentclass[]{article}
\usepackage{amsmath,amssymb}
\usepackage[a4paper]{geometry}
\usepackage{graphicx}
\usepackage{color,float}
\usepackage{subfig}
\title{ Thermodynamics of BTZ Black Holes in Gravity's Rainbow}
\author{Salwa Alsaleh}
 \date{ Department of Physics and Astronomy, King Saud University, Riyadh 11451, Saudi Arabia
 \\ \today}

\begin{document}

\maketitle

\begin{abstract}
In this paper, we  deform the thermodynamics of a BTZ black hole from rainbow 
functions in gravity's rainbow. The rainbow functions will be motivated from results 
in loop quantum gravity and Noncommutative geometry. It will be observed that the thermodynamics 
gets deformed due to these rainbow functions, indicating the existence of a remnant.  However, the Gibbs free energy does 
not get deformed due to these rainbow functions, and so the critical behaviour from Gibbs 
does not change by this deformation.This is because the deformation in the entropy cancel's out the temperature deformation.
\end{abstract}
\section{Introduction}
The 
Ho\v{r}ava-Lifshitz gravity is motivated by a deformation of the usual energy-momentum dispersion 
relation in the UV limit
\cite{hovrava2009quantum,hovrava2009spectral}. Another UV modification of general relativity also 
motivated by a deformation of the usual energy-momentum dispersion 
relation in the UV limit is the 
gravity's rainbow \cite{magueijo2002lorentz}.  It is interesting to note that  the 
deformation  of the usual energy-momentum dispersion in the UV limit  occurs 
in various approaches to quantum gravity such as   discrete spacetime
\cite{tHooft:1996ziz}, models based on string field theory \cite{Kostelecky:1988zi}, spacetime
foam \cite{amelino1998tests}, spin-network in loop quantum gravity (LQG)
\cite{Gambini:1998it} and non-commutative geometry \cite{carroll2001noncommutative}. 
The formalism  has been used to study various geometries motivated from string theory. 
In fact, the different Lifshitz scaling of space and time has been used to deform 
type IIA string theory \cite{gregory2010lifshitz}, type  IIB string theory \cite{Burda:2014jca},   AdS/CFT correspondence 
\cite{Gubser:2009cg, Ong:2011km, Alishahiha:2012iy,  dey2015interpolating},  dilaton black branes 
\cite{Goldstein:2010aw, bertoldi2009thermodynamics},cylindrical solutions \cite{Momeni:2017cvl} and dilaton black holes \cite{Zangeneh:2015uwa, Tarrio:2011de} and .
Gravity's rainbow is a more general theory that is
 motivated by deformation of the energy-momentum 
dispersion relation, similar to what motivates  Ho\v{r}ava-Lifshitz gravity. Hence both approaches are connected to same quantum gravity phenomenology. In fact, for a particular choice of rainbow functions, gravity's rainbow seem's to agree with  Ho\v{r}ava-Lifshitz gravity.
 As  what has been shown in \cite{Garattini:2014rwa}. 
The Lifshitz deformation of geometries has produced interesting results, 
and rainbow deformation has the same motivation, 
in this paper we will study the rainbow deformation 
of BTZ black holes.

In gravity's rainbow, the geometry depends on 
the energy of the  probe, and thus  probes of 
of different energy see the geometry  differently.
Thus,  a single metric is replaced  by a family of energy
dependent metrics forming a rainbow of metrics. Now the 
UV modification of the energy-momentum dispersion relation  can be expressed as 
\begin{equation}  \label{MDR}
E^2f^2(E/E_P)-p^2g^2(E/E_P)=m^2
\end{equation}
where $E_P$ is the Planck energy, $E$ is the energy at which the geometry 
is probed, and  $f(E/E_P)$ and $g(E/E_P)$
are the  rainbow functions. As the general relativity should be recovered 
in the IR limit, we have 
\begin{equation}
\lim\limits_{E/E_P\to0} f(E/E_P)=1,\qquad \lim\limits_{E/E_P\to0} g(E/E_P)=1.
\label{rainbowfunctions}
\end{equation}
Now the metric in gravity's rainbow \cite{magueijo2004gravity}
\begin{equation}  \label{rainmetric}
h(E)=\eta^{ab}e_a(E)\otimes e_b(E).
\end{equation}
So, the energy dependent   frame fields  are 
\begin{equation}
e_0(E)=\frac{1}{f(E/E_P)}\tilde{e}_0, \qquad e_i(E)=\frac{1}{g(E/E_P)}\tilde{%
	e}_i. 
\end{equation}
Here $\tilde{e}_0 $ and $ \tilde{e}_i$ are the original energy independent frame fields. 

The rainbow deformation of geometry motivated from string theory,
such as black rings \cite{Ali:2014yea}, and 
black branes \cite{ashour2016branes} has been studied. The rainbow deformation 
of higher dimensional black holes has important consequences 
for the detection of black holes at the LHC \cite{Ali:2014qra}. 
The rainbow deformation of modified theories of gravity, and 
of gravity coupled to non-linear  sources has been studied \cite{Hendi:2016dmh,hendi2016charged, Hendi:2015hja, hendi2016charged, hendi2016charged, Rudra:2016alu,Hendi:2017pld,  Garattini:2012ec}. 
The gravity's rainbow has also been used to address the information paradox in black holes 
\cite{Ali:2014cpa, Gim:gim2015black,ali2015gravitational}. It may be noted that general properties of energy dependent 
metric for a BTZ black hole, and its coupling to non-linear sources has been discussed 
using gravity rainbow \cite{hendi2016asymptotically}. In this paper, we analyse the thermodynamic aspects 
of such a deformation explicitly.  We are able to show that even though many  thermodynamic 
quantities of a  BTZ black hole are deformed by gravity's rainbow, the Gibbs free energy is not 
deformed. Thus, the critical phenomena based on Gibbs free energy is not deformed by gravity's rainbow. 
\section{BTZ Black holes}
In 2+1 dimensions, Einstein field equations with negative cosmological constant (AdS spacetime) admit - in addition to the vacuum solution- a two-parameter family of black hole solutions found by Banados, Teitelboim and Zanelli \cite{banados1992black}, given by the metric, without charge,\footnote{Planckian units is used throughout the manuscript $ k_b=c=G=\hbar =1$}
\begin{equation}
ds^2 = -N^2 dt^2 +N^{-2} dr^2 + r^2 \left( d \phi+ N^\phi dt\right) ^2.
\label{BTZmetric}
\end{equation}
Where the functions $ N^2 = \frac{(r^2 - r_+^2)(r^2 - r_-^2)}{b^2 r^2}$ and $ (N^\phi)^2 = \frac{r_+ r_-}{b r^2}$. With $ b$ is the radius of AdS and $r_{\pm}$ is obtained when the lapse function $N$ vanishes (indicating the outer and inner horizons, respectively),
\begin{equation}
r_{\pm} = \left[  b^2 \, \frac{M}{2}\left( 1\pm \sqrt{1- \frac{J^2}{b^2 M^2}}\right) \right] ^{1/2},
	\label{horizon}
\end{equation}
with $M$ and $J$ being the mass and angular momentum of the BTZ black hole respectively. They can be therefore defined in terms of $ b, r{\pm}$ accordingly:
\begin{align}
M&= \frac{r^2_+r^2_-}{b^2}, & J&= \frac{2r_+r_-}{b}.
\label{MJ}
\end{align}
In order to study the thermodynamics of BTZ black hole we first write the first law of black hole mechanics \cite{smarr1973mass}
\begin{equation}
dM= TdS+ Jd\Omega.
\label{loi1er}
\end{equation}
The angular speed  $ \Omega $ is calculated from $ g_{tt}/g_{\phi \phi}$
\begin{equation}
\Omega = \frac{J}{b^2 M}.
\label{omega}
\end{equation}
The temperature $T_0$ is calculating for the surface curvature, with the killing vector $ K = \partial_t + \Omega \partial_\phi$, \cite{birmingham2001exact,cruz2004thermal}
\begin{equation}
T_0= \frac{r^2_+-r^2_-}{2 \pi r_+}.
\label{temp0}
\end{equation}
\begin{figure}[h!]
	\centering
	\includegraphics[scale= 0.6]{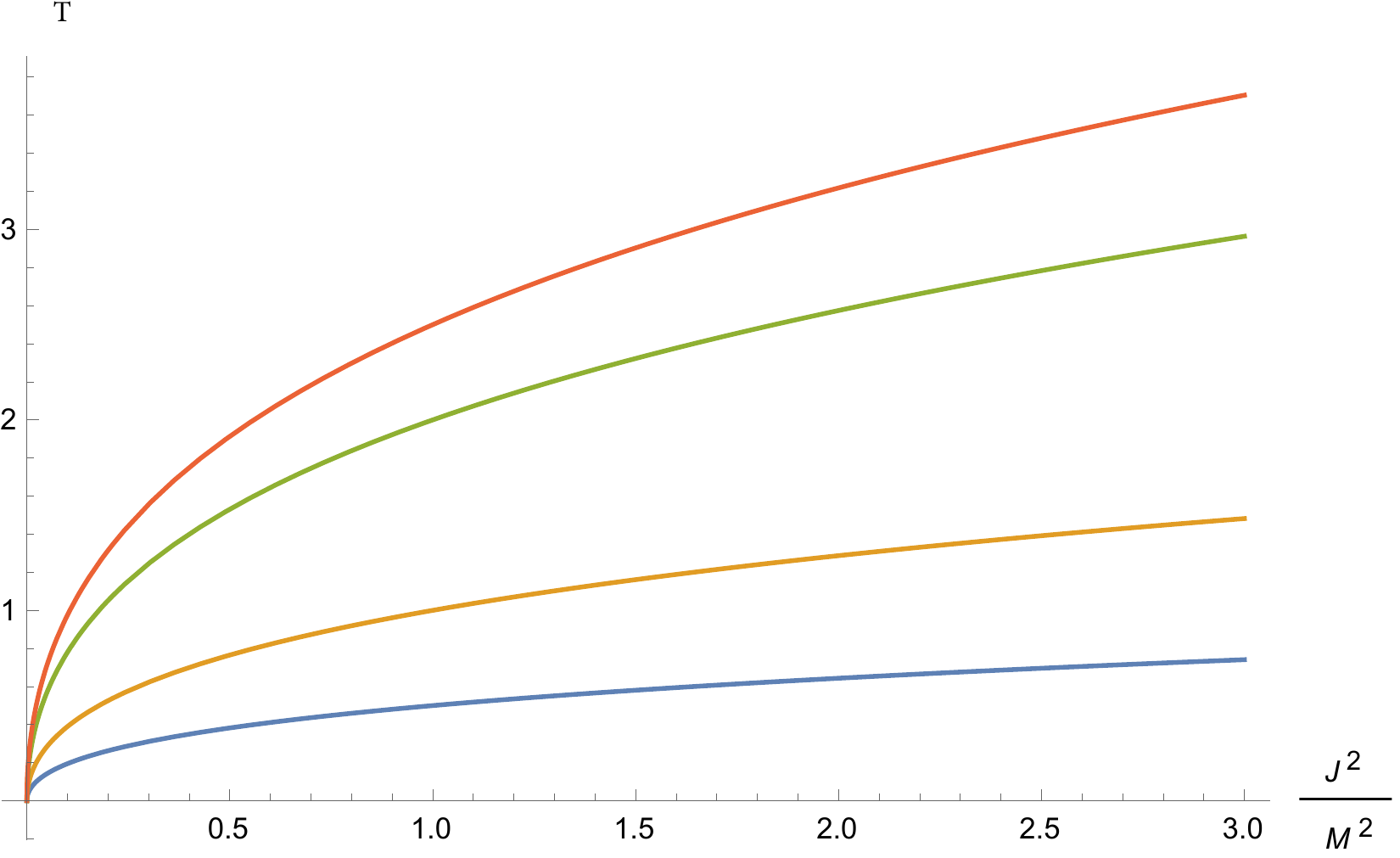}
	\caption{ The temperature of different non-extremal BTZ black holes as a function of $ \frac{J^2}{ M^2}$. For different values of $b$. Blue: $b=0.25$,yellow: $b=0.5$, green: $b=0.75$ and red: $b=1$}
	\label{tzero}
\end{figure}
The expression \eqref{temp0} indicates an extremal limit when $ M< \b J$.  We observe that the temperature of BTZ black holes show a similar thermodynamic behaviour to their higher dimensional analogues figure \ref{tzero}. Now we use \eqref{loi1er} and \eqref{MJ} to calculate the entropy
\begin{figure}[h!]
	\centering
	\includegraphics[scale= 0.6]{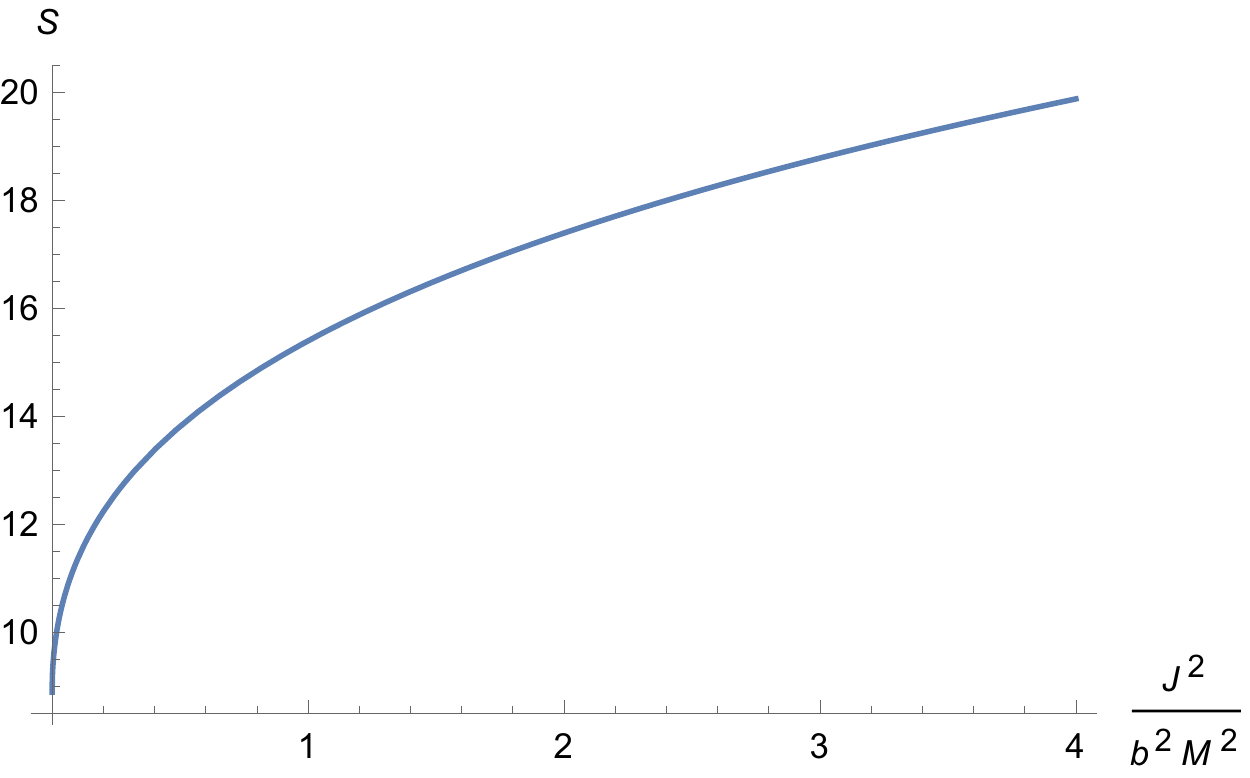}
	\caption{ The Entropy of  non-extremal BTZ black holes as a function of $ \frac{J^2}{b^2 M^2}$. Showing typical thermodynamic behaviour of Kerr-AdS black hole entropy}
	\label{szero}
\end{figure}
\begin{equation}
S_0= 4 \pi r_+ ,
\label{entropy}
\end{equation}
which is, as expected,  the forth of the horizon area `circumference'. Complying with the Bekenstein formula. We can also calculate the constant $J$ heat capacity $ C_J$ :
\begin{equation}
C_J = T\left(  \frac{\partial S}{\partial T}\right) _J = \frac{4 \pi r_+}{2-\sqrt{1-\left( \frac{J}{b M}\right) ^2}}\,\left[ \frac{1+\sqrt{1-\left( \frac{J}{b M}\right) ^2}\left( \frac{J}{b M}\right) ^2 }{2}\right] ^ {1/2}
\label{cj}
\end{equation}
Similarly, the heat capacity at constant angular velocity $C_\Omega$ is calculated,
\begin{equation}
C_\Omega = 4 \pi b \left[ \frac{M}{2}\,\left( 1+\sqrt{1-\left( \frac{J}{b M}\right) ^2}\right) \right] ^ {1/2}.
\label{comega}
\end{equation}
\begin{figure}
	\centering
	\subfloat{
		\includegraphics[scale=0.45]{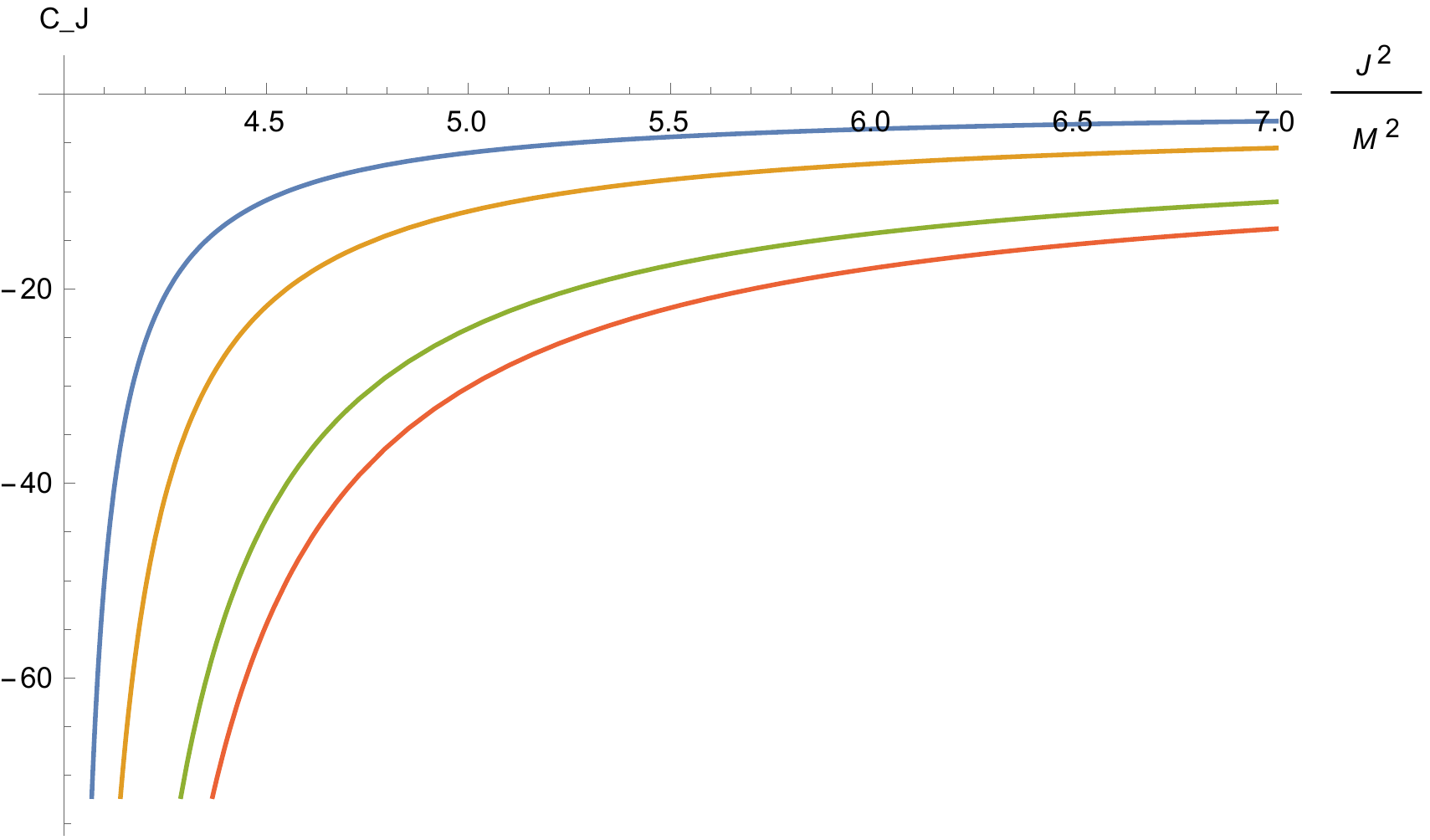}
	}
	\qquad
	\subfloat{
		\includegraphics[scale=0.45]{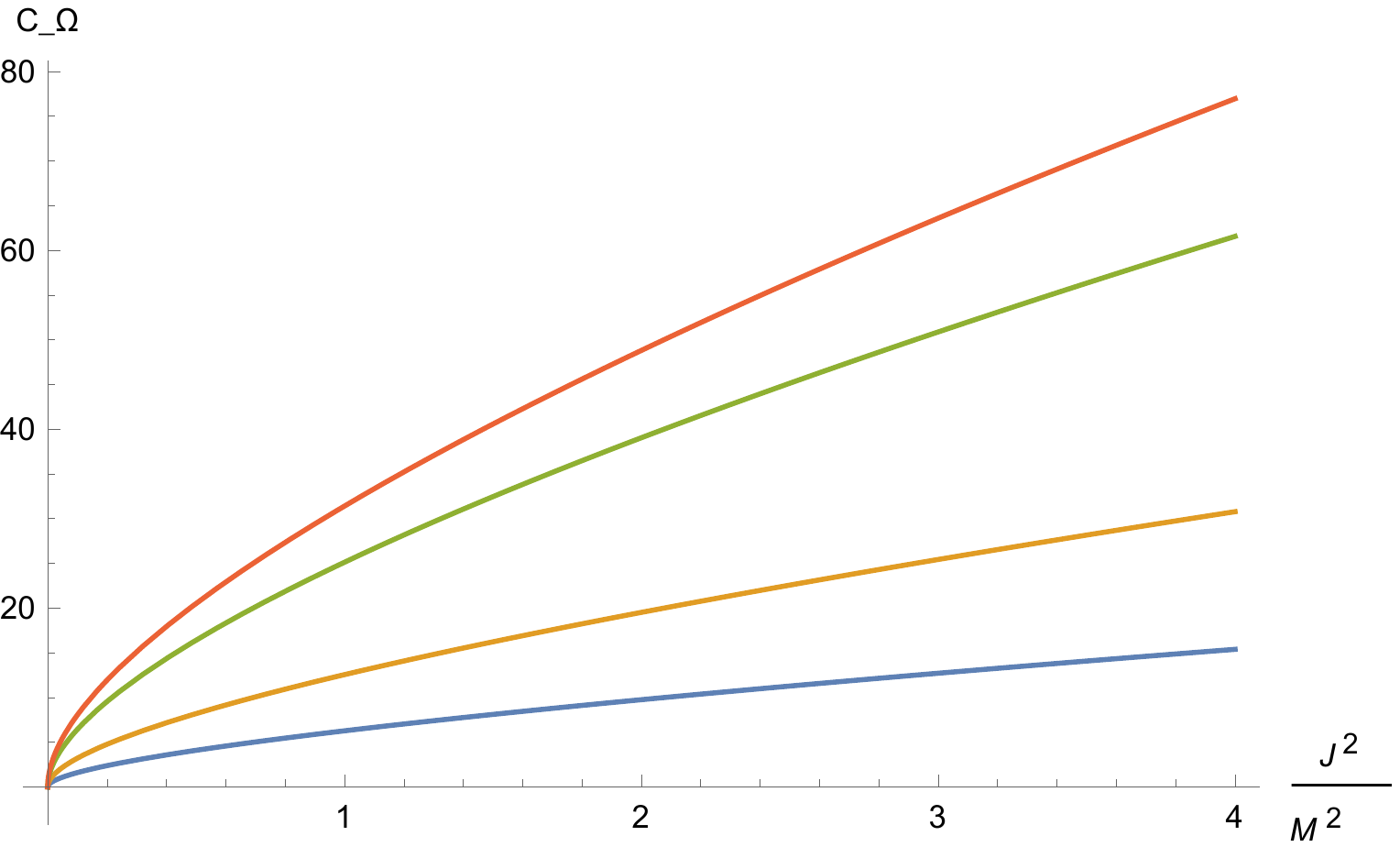}
	}
	\caption{Figures show plots of$ C_J$ and $C_\Omega$ for different BTZ black holes, as a function of $ J^2/(  M^2) $. For different values of $b$. Blue: $b=0.25$,yellow: $b=0.5$, green: $b=0.75$ and red: $b=1$}
	\label{capacity}
\end{figure}
We wish also to investigate the thermodynamic pressure - volume relation for BTZ black holes and the associated critical phenomena.  We define the ` volume' of BTZ black hole by the relation \cite{altamirano2014thermodynamics}, which is approximately the thermodynamic volume for slow rotating black holes $ J \ll1$.  
\begin{equation}
V_0= A r_+ = 16 \pi r_+^2
\label{volume}
\end{equation}
  Now, we consider the thermodynamic pressure of BTZ black hole from the Van der Wall's fluid equation of state in the extended phase space \cite{kubizvnak2012p}. 
  \begin{equation}
  P_0 := \frac{T}{v} +\mathcal{O}(J^2),
  \label{pressure}
  \end{equation}
with :
\begin{equation}
v= 2 \left( \frac{V}{\pi}\right) ^{1/2}.
\end{equation}
We observe from the PV diagram  that BTZ black holes admit the same critical phenomena as the higher dimensional Kerr-AdS black holes for some critical temperature $ T_c$ \cite{Frassino:2014pha}.
\begin{figure}[h!]
	\centering
	\includegraphics[scale= 0.6]{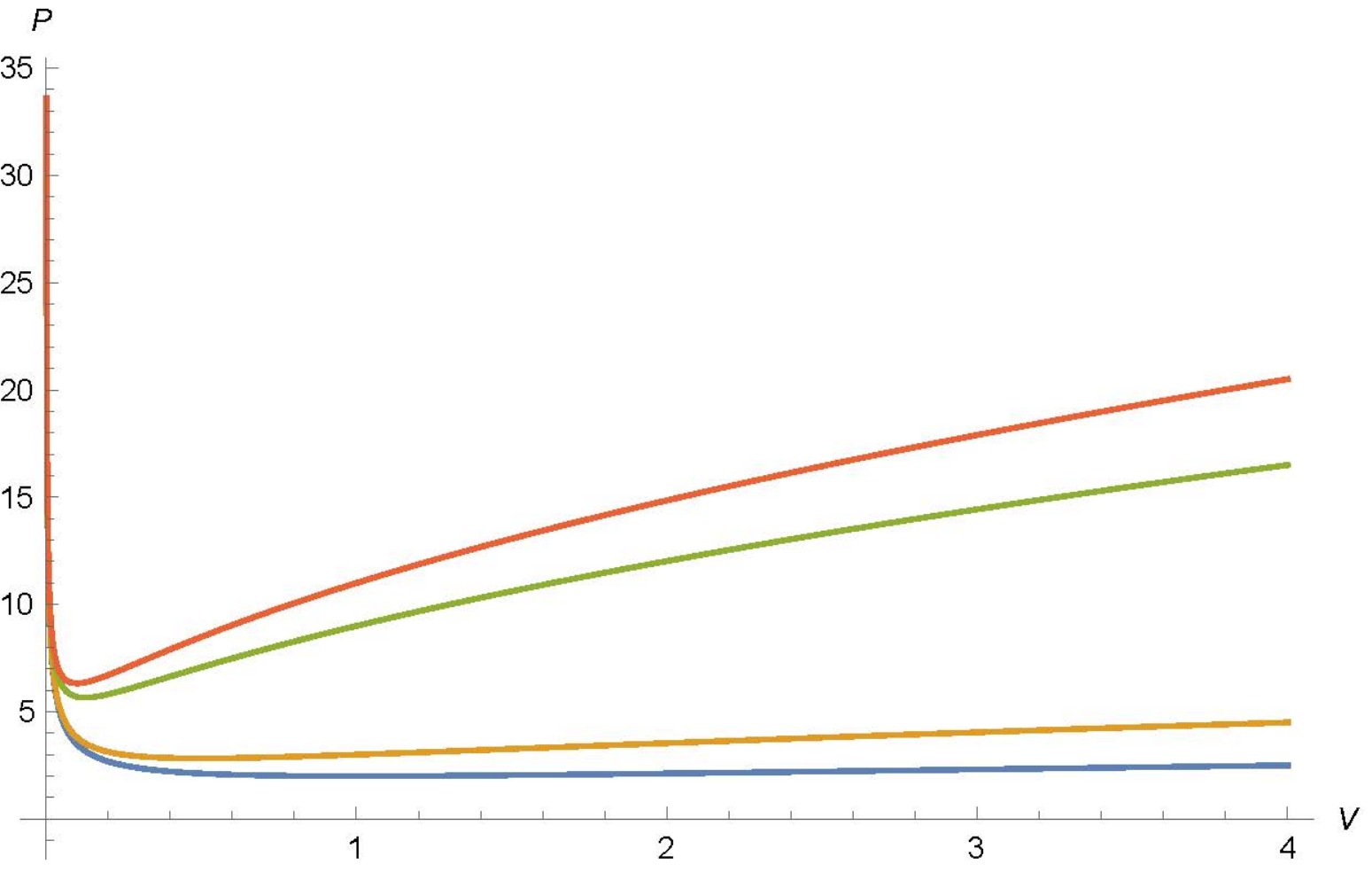}
	\caption{ PV- diagram for different BTZ black holes, two showing a critical point $ \partial P/\partial V =0$ . For different values of $b$. Blue: $b=0.25$, yellow: $b=0.5$, green: $b=0.75$ and red: $b=1$}
	\label{pv}
\end{figure}
BTZ black holes show an interesting thermodynamic properties. In the next section, they shall be studied after gravity rainbow deformation on the BTZ  metric. 
\section{BTZ black holes in gravity's rainbow}
\begin{figure}[h!]
	\centering
	\includegraphics[scale= 0.6]{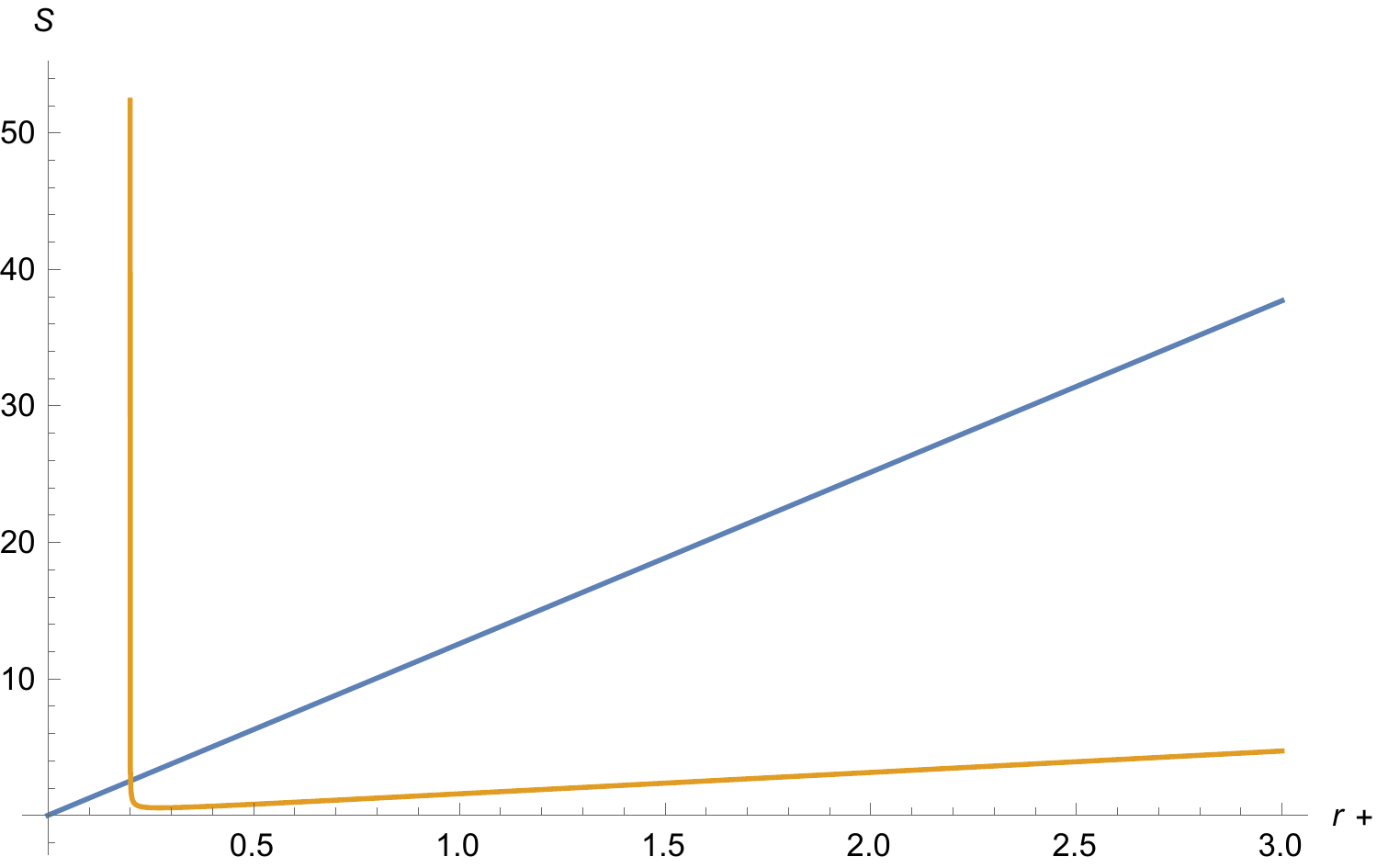}
	\caption{A plot of  BTZ entropy  $S_0$ in blue. And the deformed entropy $S$ in yellow (For $ E_p=5$, $ \eta =1$ and $ \nu=1$), both as a function of $r_+$. The latter indicates the existence of remnant of radius $\frac{1}{5}$ Planck units.   }
	\label{s_rainbow}
\end{figure}

In this section, we will deform the thermodynamics of a BTZ black hole by 
the gravity's rainbow. Now here 
$E$ is the   energy  of a 
quantum particle near the event horizon of the BTZ black hole.   This particle is  emitted from the black hole 
due to the Hawking radiation, the energy of the particle is associated with the black hole temperature $T$
\cite{ali2014black}. In fact, in the geometric units used in the paper $ k_B=1$, the black hole temperature is the same as the energy of the radiated particle, i.e $T_{BH} = E$. 
We can use the 
uncertainty principle, and write $\Delta p \geq 1/\Delta x $. Thus, we can obtain   a bound
on energy of a black hole, $E \geq 1/\Delta x $ \cite{ali2015remnant}.
This can be done for any black hole, including a BTZ black hole. It may be noted that the 
usual uncertainty principle is valid in gravity's rainbow \cite{ali2014black}.  So, 
the uncertainty in position of the particle
near the horizon of the BTZ black hole is equal to the of the 
event horizon,
\begin{equation}
E\geq 1/{\Delta x} \approx 1/{r_+}.
\end{equation}
It is important to noted from this that the energy used to deform the thermodynamics is a 
dynamical function of the radial
coordinate \cite{Garattini:2014rwa} 
The general relation for temperature of a black hole in gravity rainbow was found to be \cite{ali2015remnant} :
\begin{equation}
T= T_0 \frac{g(E)}{f(E)}
\label{Trainbow}
\end{equation}
Where $f(E)$ and $ g(E)$ are the rainbow function defined in \eqref{rainbowfunctions}. This conjecture is explained and proved in the following references \cite{ali2015gravitational,ali2015remnant,angheben2005hawking,ali2014black} and many others. We may also show that the formula \eqref{Trainbow} applies for BTZ black holes in gravity rainbow as well. \\
First, consider the modification of the metric \eqref{BTZmetric} by the gravity rainbow functions \cite{hendi2016asymptotically,ma2008hawking,ali2015gravitational}:
\begin{equation}
ds^2 = -\frac{N^2}{f^2(E)} dt^2 +g^{-2}(E)N^{-2} dr^2 + g^{-2}(E) r^2 \left( d \phi+ N^\phi dt\right) ^2.
\label{BTZrainbowmetric}
\end{equation}
An the modified temperature is given by the generic formula \cite{wald2010general}:
\begin{equation}
T = \frac{1}{4 \pi}\sqrt{\partial_r( A(r,0)) B(r,0)}
	\label{genericT}
\end{equation} 
Where $ A(r) = \frac{N^2}{f^2(E)}$ and $ B(r) = g^{-2}(E)N^{-2}$.  Therefore, we arrive at the formula for temperature in gravity rainbow \eqref{Trainbow}. \\
The are many ways in witch we can define the rainbow functions $ f(E), g(E)$. Motivated by many theoretical  \cite{garattini2014electric,awad2013nonsingular,majumder2013singularity}. And experimental \cite{amelino1998tests} approaches.  The choice of these functions in this paper is the one motivated by loop quantum gravity and non-commutative geometry \cite{amelino2013quantum,jacob2010modifications}.
\begin{align}
 f(E) &:= 1 & g(E) &:= \sqrt{1-\eta(E/E_p)^ \nu},
\label{rainbowfunctions1}
\end{align}
for $ \eta $ and $ \nu$ being free parameters.
Now, we use \eqref{Trainbow}\eqref{temp0}, and \eqref{rainbowfunctions1} to obtain the modified BTZ temperature :
\begin{equation}
T= \frac{r^2_+-r^2_-}{2 \pi r_+} \sqrt{1-\eta(1/r_+E_p)^ \nu}
\label{temprainbow}
\end{equation}
\begin{figure}[h!]
	\centering
	\includegraphics[scale= 0.6]{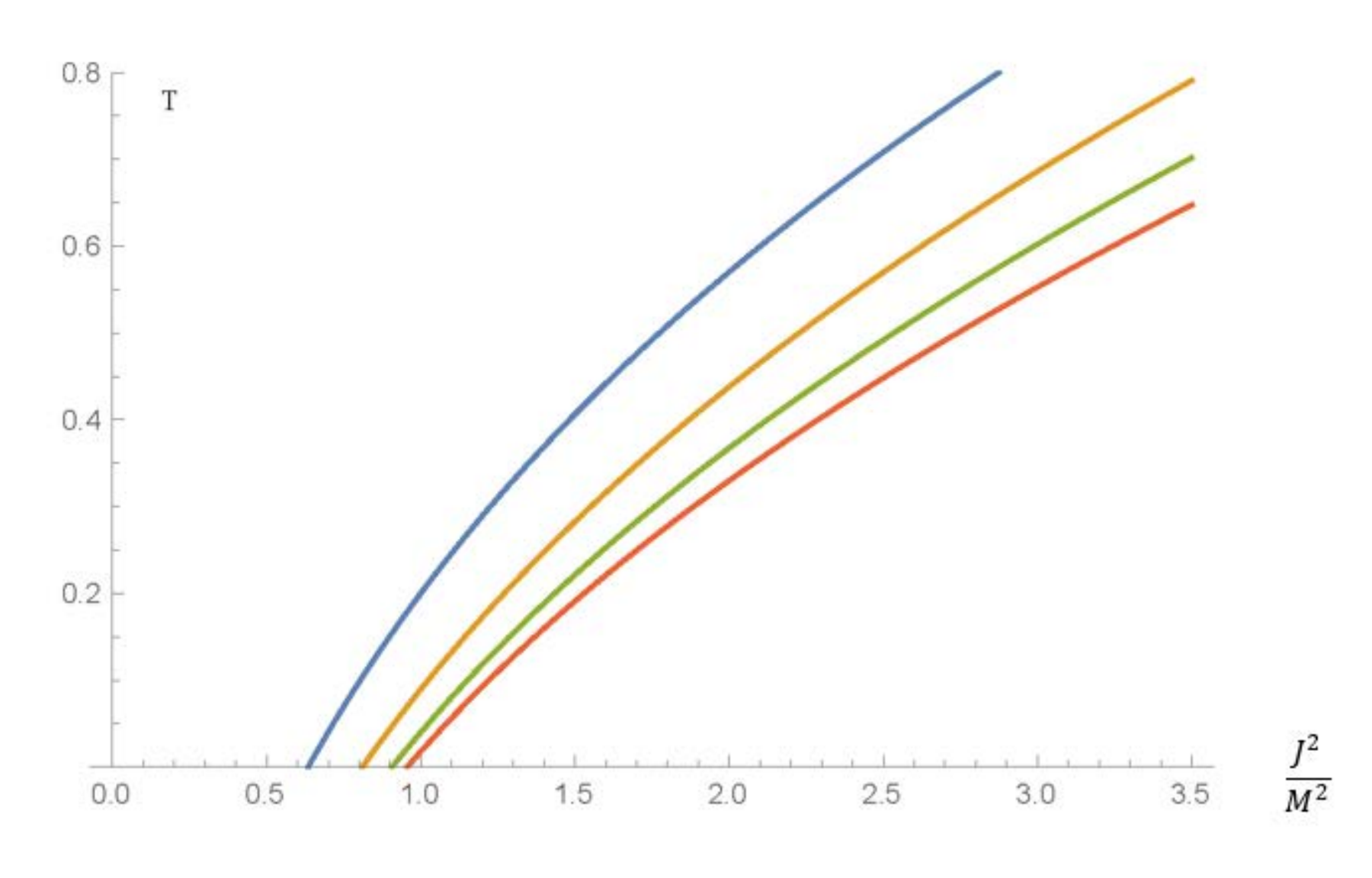}
	\caption{A plot of modified BTZ temperature as a function of $ J^2/ M^2$. For $ E_p=5$,  $ \nu= 1$ and $ \eta =1$ for different values of $b$.Blue: $b=0.25$, yellow: $b=0.5$, green: $b=0.75$ and red: $b=1$. Showing the existence of a remnant point (vanishing temperature) }
	\label{t_rainbow}
\end{figure}
In order to calculate the modified entropy, we use the first law $ dM= T dS$. With \eqref{MJ}, and \eqref{entropy} we get :
\begin{align}
S= \frac{\pi}{2} \frac{r_+}{g(E)}= \frac{\pi r_+}{2\sqrt{1-\eta(1/r_+E_p)^ \nu}}
\label{entropy_rainbow}
\end{align}
It is interesting to look at the graphs between  $ S_0$ and $r_+$, and between $S$ and $r_+$\ref{s_rainbow} . Observing that the entropy of rainbow gravity modified BTZ black holes will diminish at some point with $ r_+ \neq 0$ , indicating the existence of remnant. This effect is observed in higher dimensional Kerr-AdS black holes in rainbow gravity \cite{ali2015remnant}.
Since the entropy of a black hole is related to its area ( circumference in 2+1 dimensions). Moreover, the 2+1 dimensional gravity is a topological theory, one might relate the remnant of rainbow BTZ black hole as a topological defect associated with the minimal length of the 2+1 gravity theory \cite{vilenkin2000cosmic,carlip19952+}. The radius of this topological defect is given by:
\begin{equation}
r_{min} = \frac{\eta^{1/\nu} }{E_p} \sim \ell_p
\end{equation}
Which could be the minimal length of the 2+1 gravity rainbow. 
Now we calculate the  modified heat capacities , observe that $\left(  \frac{\partial S}{\partial T}\right) _J$ and $\left(  \frac{\partial S}{\partial T}\right) _\Omega$, do not change by the gravity rainbow modifications. Hence the heat capacities are easily computed 
\begin{align}
C_J =& \sqrt{1-\eta(1/r_+E_p)^ \nu}\left[  \frac{4 \pi r_+}{2-\sqrt{1-\left( \frac{J}{b M}\right) ^2}}\right] \,\left[ \frac{1+\sqrt{1-\left( \frac{J}{b M}\right) ^2}\left( \frac{J}{b M}\right) ^2 }{2}\right] ^ {1/2}
\label{cj_rainbow} \\
C_\Omega =&  4 \pi b \sqrt{1-\eta(1/r_+E_p)^ \nu} \left[ \frac{M}{2}\,\left( 1+\sqrt{1-\left( \frac{J}{b M}\right) ^2}\right) \right] ^ {1/2}.
\label{comega_rainbow}
\end{align}
The volume of the modified BTZ is given from \eqref{volume}:
\begin{equation}
V= \frac{2 \pi r^2_+}{\sqrt{1-\eta(1/r_+E_p)^ \nu}}\label{v_rainbow}
\end{equation}
Moreover, the modified  pressure is written as:
\begin{equation}
P= \frac{1}{2}\left( 1-\eta(1/r_+E_p)^ \nu\right) ^{1/4} P_0
\end{equation}
The P-V criticality of rainbow BTZ is not different from the ordinary BTZ black hole. This also can be seen from calculating the Gibbs free energy of ordinary and rainbow BTZ black hole,  The  Gibbs free energy, $G$ given by the thermodynamic relation:
\begin{equation}
G(T) = M+PV-TS.
\end{equation}
For an ordinary BTZ black holes it can be calculated using \eqref{MJ}\eqref{entropy} and \eqref{omega}. 
\begin{align}
G_0( T, \Omega )=& M-TS-\Omega J = -\frac{\pi ^2 b ^2 }{6\sqrt{1-\left( \frac{J}{b M}\right) ^2}}\\ \nonumber \times \qquad& \left( T^2 + \sqrt{\frac{2 \pi}{b}(1-\left( \frac{J}{b M}\right) ^2 )}T +\frac{1}{4 \pi^2 b^2} \, \left[  2+\left( \frac{J}{b M}\right) ^2 \right]  \right) \left( \frac{1}{b \Omega}\right) 
\end{align}
\begin{figure}[h!]
	\centering
	\includegraphics[scale= 0.6]{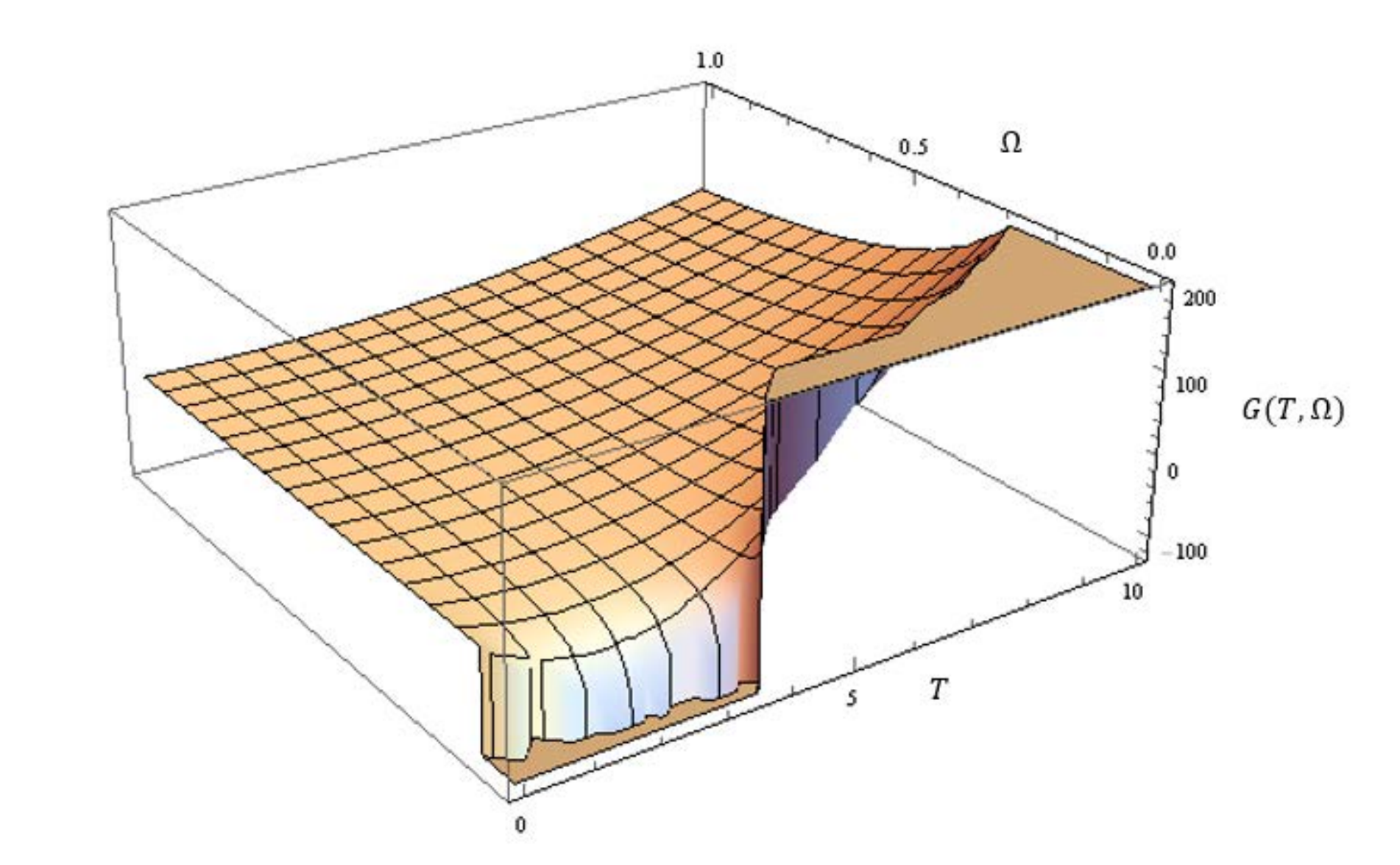}
	\caption{A plot of $G(T, \Omega)$ of a  BTZ black hole $ T>T_c$ with $b=0.5$. Showing the critical phenomena.}
	\label{Gibbs}
\end{figure}
We may conclude that $ G>0$  for critical BTZ black holes and $ G<0$ for non critical ones  \\ The Gibbs free energy for a rainbow BTZ is calculating using the same relation but, substituting $T_0$ and $S_0$ with $T$ in \eqref{Trainbow} and $S$ in \eqref{entropy_rainbow}. Since the rainbow function $ g(E)$ appear in $T$ and its reciprocal appears in $S$.  $G$ for the rainbow BTZ black hole is the same as $G_0$. Indicating the same critical phenomena. 
\section{Conclusion}
In this paper, we have deformed the geometry  of a BTZ black hole by rainbow 
functions. Thus, the thermodynamics of the BTZ black was also deformed 
by these rainbow functions. The rainbow functions that were used for this deformation 
have been    motivated from results 
in loop quantum gravity and Noncommutative geometry. It was   observed that the thermodynamics 
of the BTZ black hole got  deformed due to these rainbow functions. The graphs of the deformed entropy $S$ and temperature $T$ indicate the existence of a remnant at the last stage of evaporation, similar to the higher dimensional deformed black holes.
However, the Gibbs free energy did  
not get deformed,  and so the critical behaviour from Gibbs 
did not change by this deformation. Thus, the critical behaviour of BTZ black holes in gravity's 
rainbow was the same the  critical behaviour of BTZ black hole in ordinary gravity. This is apparent because the temperature is deformed in an opposite way to the entropy, causing both deformations to cancel out in the Gibb's free energy. 
\section*{Acknowledgements}
{ \fontfamily{times}\selectfont
	\noindent 
	Warm regards to Dr Mir Faizal for his generous help improving this work. \\ This research project was supported by a grant from the " Research Center of the Female Scientiffic and Medical Colleges ", Deanship of Scientiffic Research, King Saud University. \\
	The author would like to thank the refrees for their helpful comments that improved the paper.
\bibliography{ref}
\bibliographystyle{plain}
\end{document}